\documentclass[twocolumn,printnumbers,amsmath,amssymb,showpacs]{revtex4}
\usepackage{graphicx}
\usepackage{color}

\begin{document}
\title{Direct determination of the size of basins of attraction of jammed solids}

\date{\today}

\author{Ning Xu$^{1}$}
\author{Daan Frenkel$^{2,3}$}
\author{Andrea J. Liu$^4$}

\affiliation{$^1$CAS Key Laboratory of Soft Matter Chemistry $\&$ Department of Physics, University of Science and Technology of China, Hefei 230026, P. R. China; $^2$Department of Chemistry, University of Cambridge, Cambridge CB2 1EW, UK; $^3$FOM Institute for Atomic and Molecular Physics, Kruislaan 407, 1098 SJ Amsterdam, The Netherlands; $^4$Department of Physics and Astronomy, University of Pennsylvania, Philadelphia, PA 19104}

\begin{abstract}
We propose a free-energy based Monte-Carlo method  to  measure the volume of potential-energy basins in configuration space. Using this approach we can estimate the number of distinct potential-energy minima, even when this number is much too large to be sampled directly.   We validate our approach by comparing our results with the direct enumeration of distinct jammed states in small packings of frictionless spheres.  We find that the entropy of distinct packings is extensive and that the entropy of distinct hard-sphere packings must have a maximum as a function of packing fraction.
\end{abstract}


\maketitle
When many equal-sized spheres are poured into a container, the spheres are unlikely to end up arranged in a periodic lattice. This observation reflects the fact that $S_0$, the entropy of distinct disordered packings that are mechanically stable, is very large compared to the corresponding entropy of distinct ordered packings.  The fact that $S_0$ is so large has important consequences for the disordered packings such as granular materials~\cite{edwards,song,zamponiRMP}.

There is a natural connection between hard-sphere packings and glasses~\cite{ohern}, whose potential energy landscapes have many minima (inherent structures~\cite{stillinger}), corresponding to mechanically-stable states.  These minima have been argued to be relevant for our understanding of the glass transition~\cite{stillinger2,zamponiRMP}.  The number of such minima has been calculated from replica theory~\cite{zamponi,zamponiRMP,zamponisoft}.  In calculating this number numerically, however, a protocol must always be used to generate energy minima.  Typical protocols produce states with probabilities that are not known; for example, when the entropy of minima is calculated from finite-temperature simulations~\cite{sciortino,sastry1,doliwa,massen}, one must assume that the temperature is low enough so that the system crosses no barriers.  Similarly, when the entropy is calculated from algorithms that involve compression or dilation of the system~\cite{lubachevsky,xu,jiao}, it may depend--even under ideal conditions--on the algorithm used. As a result, it is difficult to count the number of distinct mechanically-stable states, with states weighted equally.

In this letter we report a general computational method to measure the volume of a basin of attraction associated with an arbitrary potential energy minimum.  This is the key to calculating the entropy of distinct minima for soft spheres because there is a protocol that generates minima weighted by their basin volumes~\cite{ohern}.  In this ``basin" protocol, states in configurational space are selected at random and each one is quenched to its nearest energy minimum~\cite{ohern}.  By using this protocol and correcting for the weighting by calculating the basin volume, we can obtain the unweighted entropy of distinct mechanically-stable states (packings) for soft spheres.  Finally, the analogous entropy for hard-sphere packings can be obtained from the density of soft-sphere packings at zero pressure.  We find that there must be a maximum in the entropy of distinct hard-sphere packings, at least for small systems, in agreement with earlier results obtained by direct enumeration~\cite{xu}.

To explain our approach, we first define the volume of a basin for a packing of soft spheres as
\begin{equation}
v_b=\int {\rm d}\vec{R} G(\vec{R},\vec{R}_0),
\end{equation}
where $G(\vec{R},\vec{R}_0)=1$ if, upon energy minimization, any point $\vec{R}$ in configuration space ends up at $\vec{R}_0$, the position of the local potential energy minimum, and $0$ otherwise.  The integral is over the whole configuration space.  We view the (hyper) volume associated with a given basin as a partition function and hence compute its value by a suitable free-energy calculation method. Here, we will use the standard ``Einstein'' method~\cite{FrenkelSmit} and compute the basin free energy by comparing it to the free energy of a system confined near the minimum $\vec{R}_0$ by a harmonic potential with spring constant $k$. For arbitrary $k$,  the canonical partition function of the system is:
\begin{equation}
Q(k) = \int {\rm d}\vec{R} G(\vec{R}, \vec{R}_0) {\rm exp} \left( -\beta k u^2 / 2 \right), \label{partition}
\end{equation}
where $u=|\vec{R}-\vec{R}_0|$ is the distance between $\vec{R}$ and $\vec{R}_0$, and $\beta\equiv(k_BT)^{-1}$ with $k_B$ the Boltzmann constant.  $G(\vec{R},\vec{R}_0)$ in Eq.~(\ref{partition}) can be rewritten as $\exp(-\beta U)$, where $U=0$ when $\vec{R}$ is in the basin, and $\infty$ otherwise.  Obviously, $v_b=Q(0)$.  Without loss of generality, we choose $\beta=1$.

The free energy of this system is $F(k)=-{\rm ln}Q(k)$ and
$\frac{{\rm d}F(k)}{{\rm d}k}=\langle  u^2/2\rangle _k,$
where $\langle  ...\rangle _k$ denotes a canonical ensemble average at the spring constant $k$. This average can be sampled in a standard Monte-Carlo (MC) simulation. The change in free energy upon switching on a spring constant $k_m$ is
\begin{equation}\label{eqn:FEinstein}
F(k_m) = F(0) + \int_0^{k_m} \langle u^2/2\rangle_k {\rm d}k ,
\end{equation}
where $k_m$ is chosen sufficiently large that the confining potential has no influence. In that case $F(k_m)$ is known analytically and Eq.~(\ref{eqn:FEinstein}) allows us to compute $F(0)$ and from that the volume of the basin, as  $v_b(\vec{R}_0)=\exp(-F(0))$. In practice, we  choose a maximum $k_m$ such that most (in our case $>90\%$) of the associated Gaussian distribution is within basin $\vec{R}_0$.  One then corrects the Einstein crystal result for the confining effect of the basin: $F(k_m)= -\frac{dN}{2} {\rm ln} \left( 2\pi / k_m\right) - {\rm ln} f $,
where $d$ is the dimension of space, $N$ is the number of particles in the system, and $f$ is the fraction of the associated Gaussian distribution within basin $\vec{R}_0$.

Given the basin volume, we can calculate the entropy of distinct mechanically-stable minima.  We include in our analysis only energy minima that are mechanically-stable (jammed).  The fraction of the total configuration space, $V_{tot}$, occupied by basins of jammed states, $f_j$, is computed~\cite{ohern} by quenching randomly selected points in configuration space to the nearest energy minimum and calculating the fraction that end up in jammed states.  The volume of configuration space at packing fraction $\phi$ occupied by jammed basins is
$V_c (\phi) = f_j (\phi) V_{tot}$.

As pointed out by Speedy in a slightly different context~\cite{speedy}, the total configuration space can be uniquely decomposed into distinct basins and hence its volume is simply the sum of the volumes of the constituent basins. Thus,
\begin{equation}
V_c(\phi)=\sum_{i=1}^{\Omega_c} v_b = \Omega_c \left(\frac{1}{\Omega_c}\sum_{i=1}^{\Omega_c} v_{b,i} \right)= \Omega_c \left<v_b\right>. \label{volume}
\end{equation}
By {\em sampling} the basin volume to obtain $\left<v_b\right>$, we can therefore compute $\Omega_c$, the total number of distinct jammed states.

Note that ``computing the average basin volume'' sounds simpler than it is because the probability to sample a given basin is proportional to the basin volume itself. We correct for this bias by dividing by the basin volume. However, if a substantial fraction of all distinct basins together occupy a negligible volume of configuration space, they will not be sampled at all. For this reason, it is imperative to check this method for small systems for which all distinct basins can be identified.

\begin{figure}
\vspace{0.in}
\includegraphics[width=0.45\textwidth]{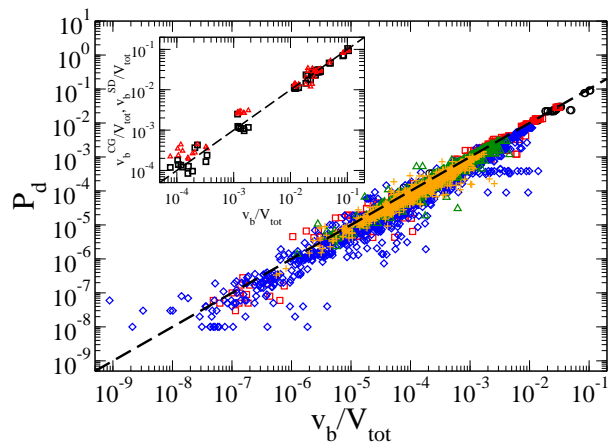}
\caption{\label{fig:check} (color online) Probability of finding a given minimum calculated in two ways:  from direct enumeration, $P_d$, and from MC calculations of the basin volume relative to the total volume of configuration space, $v_b/V_{tot}$.  Included are systems at packing fraction $\phi=0.9$ of $N=8$ (black circles), $10$ (red squares), $12$ (blue diamonds), $14$ (green upward triangles), and $16$ (orange pluses) particles.  For $N=8$, $10$, and $12$, all distinct states are shown, while for $N=14$ and $16$ only the first $1000$ states are shown.  The dashed black line is $P_d=v_b/V_{tot}$.  Inset: the volumes of all distinct basins for $N=8$, calculated by steepest descent ($v_b^{SD}$, red triangles) and conjugate gradient ($v_b^{CG}$, black squares) compared to volumes $v_b$ calculated by the L-BFGS algorithm.  The dashed black line is $v_b^{CG(SD)}=v_b$.
}
\end{figure}

To test the method, we consider $N$ disks in a square box of length $L$ with periodic boundary conditions.  Disks $i$ and $j$ interact via a ``harmonic'' repulsion   $V_{ij}=\epsilon\left( 1 - r_{ij} / \sigma_{ij} \right)^2/2$ when the distance between their centers of mass, $r_{ij}$ is smaller than the sum of their radii, $\sigma_{ij}=\left(\sigma_i + \sigma_j \right)/2$, and zero otherwise.  In order to avoid crystallization, we use a $50:50$ binary mixture of disks.  The diameter ratio of the large disks to the small ones is $1.4$.  We choose units where the length of the simulation box is $L=1$ and the characteristic energy of the interaction is $\epsilon=1$.  For this system, the total volume of configuration space occupied by distinct basins is
\begin{equation}
V_{tot}=\frac{L^{dN}}{\left[ \left( N/2\right)!\right]^2} \label{Vtotdef}
\end{equation}
where $\left[ \left( N/2\right)!\right]^2$ accounts for disk indistinguishability.

The direct calculation of the integral on the right hand side of Eq.~(\ref{eqn:FEinstein}) is computationally expensive because the acceptance step of every MC move requires an energy minimization (to see if the system has left the original basin). Otherwise, the calculations are exactly as in Ref.~\cite{FrenkelSmit}. In what follows, we use Gauss-Lobatto quadrature to evaluate Eq.~(\ref{eqn:FEinstein}), changing variables so that the integrand varies only weakly over the integration interval to improve accuracy (see~\cite{FrenkelSmit}).  We verified that the integrand in the Gauss-Lobatto quadrature indeed varies smoothly with increasing force constant $k$ of the harmonic spring.

As the potential energy has to be minimized at every step, the efficiency of the energy minimization routine becomes important. From any given starting point, the routine should find the minimum corresponding to a steepest-descent (SD) search. Only the SD algorithm itself is guaranteed to do that, but this algorithm is not efficient at finding the minimum. In what follows, we make use of the L-BFGS minimization routine \cite{lbfgs} as it is much (an order of magnitude) faster. We find that the L-BFGS, conjugate gradient (CG) and SD algorithms yield very similar results for basin volumes and volume distributions for $N=8$, as shown in the inset to Fig.~{\ref{fig:check}}. However, in general it may be safer to use SD, in spite of its higher computational cost.

The first step in the computation of a basin volume is to find a potential energy minimum.  To do this, we generate a random point in the configuration space of the system under study (a $dN$-dimensional hypercube for a system of $N$ spheres in $d$ spatial dimensions).
Starting from this initial coordinate, the potential energy of the system is minimized to find the coordinate $\vec{R}_0$ that corresponds to the (local) potential minimum \cite{ohern}. Since the probability of sampling a given minimum is proportional to the volume of its ``catchment basin", we can deduce the volumes of the individual basins from the frequency with which they are sampled, for systems sufficiently small so that all basins can be sampled in a simulation.  Thus, this brute-force approach can be used to validate the free-energy based volume calculation.

\begin{figure}
\vspace{0.in}
\includegraphics[width=0.45\textwidth]{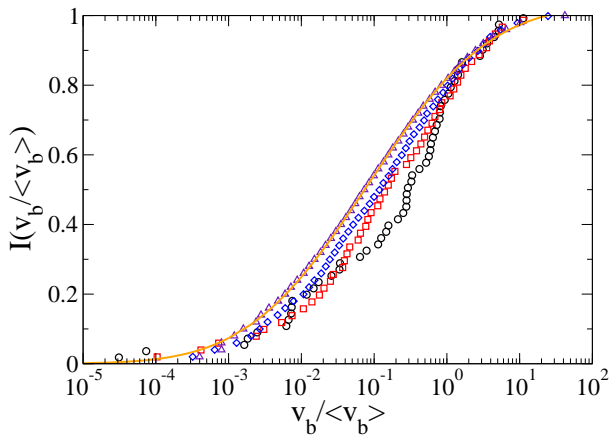}
\caption{\label{fig:disv} (color online) Cumulative distribution of the basin volume, $I(\frac{v}{\langle  v\rangle })$ for binary mixtures with $N=10$ (black circles), $12$ (red squares), $14$ (blue diamonds), and $16$ (purple triangles), all at $\phi=0.9$.  The orange solid curve shows the quasi log-normal fit to the $N=16$ data according to Eq.~(\ref{dis}) with $a=0.23$, $b=0.60$, and $c=1.04$.
}
\end{figure}

We used the two approaches mentioned above to compute the number of distinct catchment basins, $\Omega_c$, of the binary disk mixture at a packing fraction $\phi=0.9$ and system sizes $N\in[8, 16]$. For these small systems, we can find effectively all distinct states by sampling up to $N_t=10^{8}$ uncorrelated initial configurations~\cite{xu}. During the runs, we keep track of $n_s(n_t)$, the number of distinct basins sampled after $n_t$ randomly chosen initial configurations.  As shown in Ref.~\cite{xu}, $n_s$ saturates for large $n_t$, suggesting that we have found all distinct basins or, more precisely: the combined volume of all basins not sampled is less than ${\mathcal O}(n_t^{-1})$.    The fractional volume occupied by an individual basin $i$ is then given by $P_d(i)\equiv n(i)/N_t$, where $n(i)$ denotes the number of times that we have sampled the same basin $i$ after $N_t$ trials.

For each distinct basin, we also calculate the basin volume $v_b(i)$ using the free-energy method described above.   The fractional volume occupied by distinct basin $i$ is given by $v_b(i)/V_{tot}$, where $V_{tot}$ is given by Eq.~(\ref{Vtotdef}).

Fig.~\ref{fig:check} shows the correlation between  $P_d$ and $v_b/V_{tot}$ for each of the distinct basins $i$ obtained from the direct enumeration.  The dashed line is not a fit but corresponds to the relation $P_d=v_b/V_{tot}$.  Thus, Fig.~\ref{fig:check} shows that the free-energy calculation of the basin volumes works very well, even though the shapes of the high-dimensional basins are very complicated.

\begin{figure}
\includegraphics[width=0.45\textwidth]{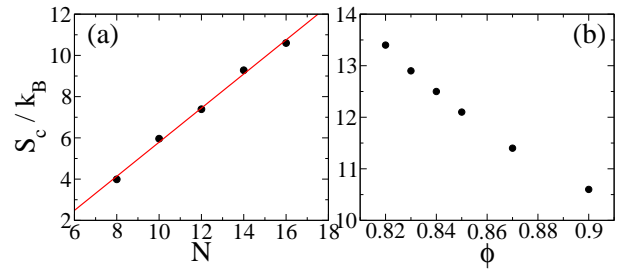}
\caption{\label{fig:entropy}  (color online) Configurational entropy $S_c/k_B$ as a function (a) system size $N$ at $\phi=0.9$, and (b) packing fraction $\phi$ at $N=16$.  The red line in (a) is the linear fit to the data: $S_c=0.83N-2.48$.
}
\end{figure}

It is straightforward to calculate the average basin volume $\langle v_b\rangle$ when all the distinct basins are known.  But for larger systems for which only a small subset of basins can be identified,  $\langle v_b\rangle$ can be calculated only if the {\it distribution} of basin volumes, $P(v_b)$, scales in a known fashion with system size.  Fig.~\ref{fig:disv} shows the cumulative distribution of the basin volumes  ($I(\frac{v_b}{\langle v_b\rangle})$).  For larger $N$, the cumulative distribution $P(v_b)$ is well represented by
\begin{equation}
I\left(x\right) = \frac{\{{\rm erf} \left[ a {\rm ln}\left( x\right) + b\right] + 1\}^c}{2}, \label{dis}
\end{equation}
where $x=\frac{v_b}{\langle v_b\rangle}$, while  $a$, $b$, and $c$ are adjustable parameters.  As $N$ increases, $a$ and $c$  decrease slightly, while $b$ increases.  Specifically, $c \rightarrow 1$, suggesting that the distribution $P(v_b)$becomes log-normal for larger systems.

This result is perhaps not surprising if one expects the distribution of the entropy of states within a basin (the logarithm of the basin volume) to be Gaussian in the thermodynamic limit.  If this is indeed the case, then one can compute the average basin volume (and hence the total number of distinct basins) from a simulation that samples only a fraction of all basins.

Once $I(v_b)$ has been obtained for a given system size, the configurational entropy $S_c$  follows, using Eq.~(\ref{volume}).  Fig.~\ref{fig:entropy}(a) shows that the configurational entropy $S_c = k_B \ln \Omega_c$ is extensive, i.e. it scales linearly with $N$.  This is expected for large systems~\cite{stillinger,stillinger1} but not necessarily for the sizes studied here.  Fig.~\ref{fig:entropy}(b) shows the variation of $S_c$ with the packing fraction $\phi$. The number of distinct states increases as $\phi$ decreases, as expected.  Note that $S_c(\phi)$ is the configurational entropy of distinct jammed energy minima in soft sphere packings, or equivalently, the entropy of distinct mechanically-stable packings.

\begin{figure}
\vspace{0.in}
\includegraphics[width=0.45\textwidth]{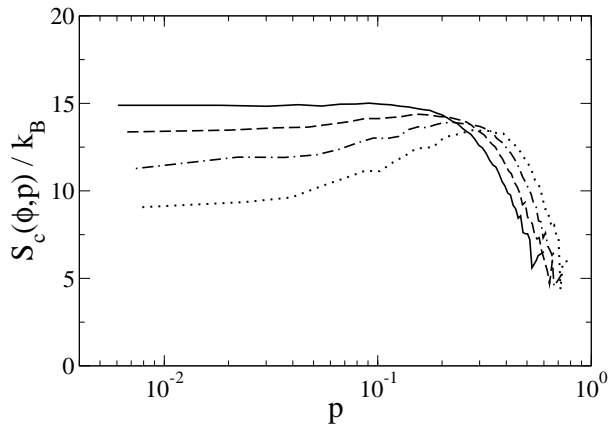}
\caption{\label{fig:Scp} Entropy of distinct minima at packing fraction $\phi$ and pressure $p$, $S_c(\phi,p)/k_B=\ln \Omega(\phi,p)$, where $\Omega(\phi,p)$ is the density of states between $p$ and $p+dp$.  Here, $S_c(\phi,p)$ is shown for systems of $N=16$ particles at $\phi=0.82$ (solid), $0.83$ (dashed), $0.84$ (dot-dashed) and $0.85$ (dotted).  $S_c$ reaches a well-defined value as $p \rightarrow 0$.
}
\end{figure}

However, the entropy of distinct mechanically-stable packings of {\it hard} spheres, $S_0(\phi)$, is not the same as that for soft spheres $S_c(\phi)$.  To obtain the former quantity, we must look only at soft-sphere packings that are at the jamming threshold ({\it i.e.} at zero pressure, $p=0$) at each packing fraction~\cite{ohern}.  Fortunately, we can calculate this directly from the sampled soft-sphere minima without introducing a protocol for bringing the system to $p=0$ that might bias the weightings of the resulting states~\cite{xu}.  We calculate the distribution $P(\phi,p)$ of basins whose minima have pressure $p$ at $\phi$ and use the average basin volume, $\langle v_b \rangle$, to obtain the density of states of distinct energy minima, $\Omega(\phi,p)$, with pressures between $p$ and $p+dp$, via Eq.~(\ref{volume}).  The entropy of distinct jammed hard-sphere packings is then $S_0(\phi)=S_c(\phi,p=0)=k_B \ln \Omega(\phi,p=0)$.

Fig.~\ref{fig:Scp} shows that $S_0(\phi)$ increases with decreasing $\phi$ over the range studied.  However, we also know that $S_0$ must vanish at sufficiently small $\phi$.  Thus, $S_0$ must have a maximum, in agreement with earlier estimates~\cite{xu} and theoretical predictions~\cite{zamponisoft}.  It would be interesting to explore the connection between this maximum and the random close-packing density in large systems~\cite{kamien}.

In summary, the free-energy method proposed here allows us to compute the volume of individual basins in the energy landscape of a many-particle system. This, in itself, is an extremely useful result.  We also find that from the distribution of basin volumes we can obtain the number of distinct energy minima (the number of distinct jammed packings).  Here, we have tested our method for small systems where all basins can be identified by brute force, but our method can be applied to far larger systems, where direct enumeration is impossible.  In practice, the reliability of this approach depends strongly on the existence of a universal form for the functional form of the distribution basin volumes. Further tests are needed, but our results suggest that a log-normal form may be appropriate for larger system sizes.

We thank S. R. Nagel for his contributions to this work, and P. M. Chaikin and F. Zamponi for stimulating discussions.  This work is supported by NSFC-11074228 (NX), ERC Grant 227758 (DF), EPSRC RG58958 (DF), Wolfson Merit award RG50412 (DF), and by the DOE Office of Basic Energy Sciences through DE-FG02-05ER46199 (AJL, NX) and DE-FG02-03ER46088 (NX).

\end{document}